# CasGCN: Predicting future cascade growth based on information diffusion graph

Zhixuan Xu[1], Minghui Qian[1], Xiaowei Huang[2], Jie Meng[3]
[1]School of Information Resource Management, Renmin University of China, Beijing, 100872 China
[2]Department of Computer Science, University of Liverpool, Liverpool, L69 3BX United Kingdom
[3]Institute for Digital Technologies, Loughborough University, London, E20 3BS United Kingdom

Corresponding author: Zhixuan Xu (e-mail: xzx@ ruc.edu.cn).

**ABSTRACT** Sudden bursts of information cascades can lead to unexpected consequences such as extreme opinions, changes in fashion trends, and uncontrollable spread of rumors. It has become an important problem on how to effectively predict a cascade' size in the future, especially for large-scale cascades on social media platforms such as Twitter and Weibo. However, existing methods are insufficient in dealing with this challenging prediction problem. Conventional methods heavily rely on either hand crafted features or unrealistic assumptions. End-to-end deep learning models, such as recurrent neural networks, are not suitable to work with graphical inputs directly and cannot handle structural information that is embedded in the cascade graphs. In this paper, we propose a novel deep learning architecture for cascade growth prediction, called CasGCN, which employs the graph convolutional network to extract structural features from a graphical input, followed by the application of the attention mechanism on both the extracted features and the temporal information before conducting cascade size prediction. We conduct experiments on two real-world cascade growth prediction scenarios (i.e., retweet popularity on Sina Weibo and academic paper citations on DBLP), with the experimental results showing that CasGCN enjoys a superior performance over several baseline methods, particularly when the cascades are of large scale.

## I. INTRODUCTION

The cascade of information diffusion is a pervasive phenomenon in Weibo, Tweeter, Facebook, and other social network platforms. Once a user retweets a micro-blog, a photo, or a video from another person, some of his/her followers who see it may also share it with their followers, then a cascade of information diffusion forms. Although most information cascades' size is small [1], once a super cascade occurs, it can make a huge influence on the internet [2]. For example, a popular praise of a product can bring considerable benefits to the marketers and companies. In contrast, negative information cascades -- such as the disinformation or misinformation -- can lead to catastrophic consequences. By these reasons, predicting the future popularity (or more precisely, the final retweet size of a cascade in a social network) is greatly valuable for managers to make strategic decisions or take precautions before the unexpected results [3]. However, since the social network is generally a large-scale open system and the mechanism of cascade evolution is still inconclusive, the cascade growth prediction is still a challenging problem.

The current cascade growth prediction methods share one common rationale: to capture predictive features in the early state of cascades and model the relationship between these features and cascades' future popularity. Existing methods can be roughly divided into three categories, namely feature-based approaches, generative approaches, and graph representation approaches. Feature-based approaches heavily rely on a set of well-designed features that are mainly hand-crafted and sometimes difficult to be generalized [2,5,6,7]. Generative approaches simulate the future retweet dynamics [8,9,10], by having strong assumptions that usually oversimplify the real diffusion process. For instance, Shen et al. [8] assume the uniform contribution of each new retweet for future cascade growth. However, in reality, the opinion leader with a significant number of followers often brings more retweets than normal users.

Graph representation approaches attempt to predict cascade popularity through end-to-end deep learning models, which embed an early cascade graph into a lower-dimensional space and then make predictions based on the obtained embeddings [3,4,14]. Current graph representation approaches are based on the standard neural network like recurrent neural networks (RNNs), which cannot directly work with graphic inputs. As a result, they have to rely on techniques -- such as the random walk [3] or skip-gram [14] -- to convert the graphs to regular Euclidean data. Unfortunately, such conversion significantly increases the computation redundancy. Additionally, as they cannot explain what features the model exactly captures from data generated by random walks, the end-to-end approaches are often not interpretable. Moreover, the obtained graph embeddings are not compatible with the node's temporal features, making the models hard to be generalized to different application scenarios. For example, Cao et al. [4] split the observation time window into several time intervals and learn each interval's time decay effect. However, since the time effect in different types of cascades is not always the same, it is challenging to set an appropriate length of time interval for all application scenarios.

There is still a room for improvement of current cascade prediction methods. Given the recent progress of graph learning techniques [15,16,17], the graph convolutional networks (GCNs) inspire us to investigate a cascade graph convolutional learning framework to overcome all those deficiencies mentioned above. Fig.1 presents an illustrative diagram of our proposed graph convolutional framework. Specifically, the framework directly uses cascade graphs as input without any data conversion processes to improve the computation efficiency. The structural features of each node are extracted by a carefully designed graph convolutional layer which can not only capture connection relationships between nodes but also distinguish different information diffusion directions. The obtained embeddings containing nodes' structure information are also compatible with the other node-level features (e.g., temporal features) and thus more flexible for different application scenarios. The final predictions are made based on two critical factors that have been proved by previous studies, i.e., the node's structural features and temporal features, in order to make the model easier to be interpreted.

Although GCNs have shown strong power of handling non-Euclidean data like chemical molecular structure [18], knowledge graphs [19], and many other research areas. There are still challenges to build such a framework for cascade growth prediction. Unlike the undirected graphs in those areas where GCNs are successfully applied, the cascade graph is usually a directed graph. Therefore, the first challenge is to consider a node's structural feature from both in and out directions in a cascade graph. The second challenge is to aggregate all nodes' structural features and temporal features for final cascade predictions.

We conduct experiments on two real-world information cascade phenomenons, one is to predict retweet popularity of messages on the social platform, and another one is to predict citation sizes of academic papers. We compare the prediction performance of our method with that of several baseline methods. The results show that the CasGCN outperforms not only the feature-based models but also the state-of-art cascade prediction models based on deep learning methods. Besides, to verify the effectiveness of CasGCN's components, we develop several variants of the proposed model. Compared with the original CasGCN, all the modified models show different performance degradations.

The contribution of our work is as follows:

(1) We propose a novel graph convolutional learning framework called CasGCN which directly uses cascades' topological graphs as input and combines nodes' structural features with nodes' temporal features to predict the future growth sizes of information cascades;

(2) As the information diffusion direction provides valuable information of a cascade's structure, we develop a bio-directional convolution technique to aggregate the information of a node's neighbours from both directions;

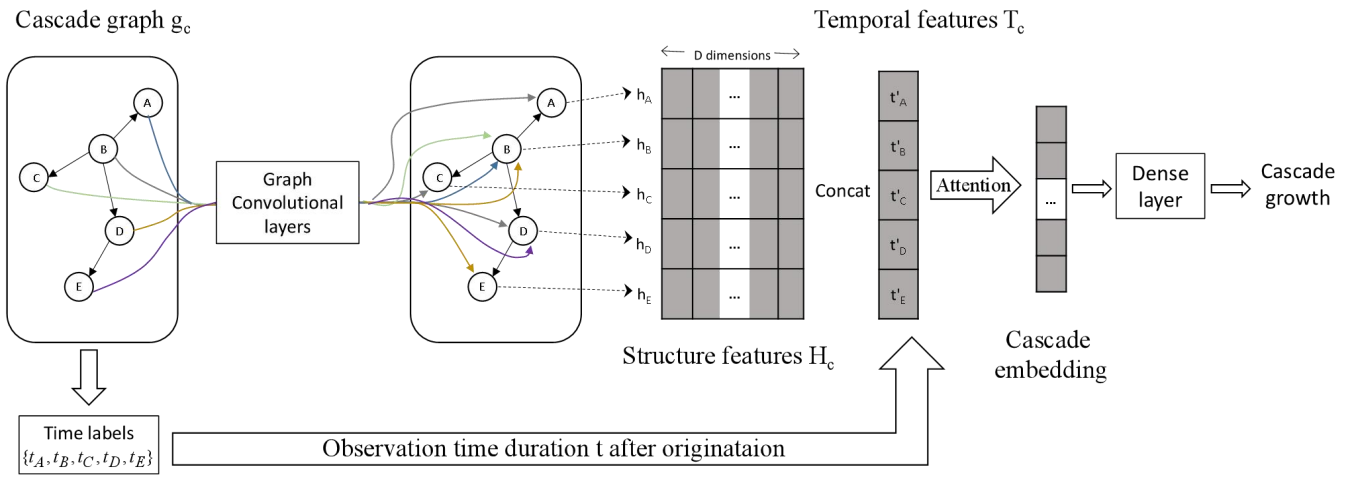

**FIGURE 1.** The framework of CasConv model

(3) We also generate several datasets with fewer small-scale cascades to reduce the biases caused by imbalanced data, and we apply the CasGCN as well as other baselines on these datasets to better illustrate the models' performances on large-scale cascades;

(4) We make performance comparisons between our proposed model and extensive cascade prediction models, demonstrating that the CasGCN significantly outperforms the feature-based models, the state-of-art deep graph representation models and the node embedding model.

In the rest of paper, Section 2 makes a literature review of related work. In Section 3 we define the problems and provide a specific illustration of our method. The results of performance comparisons and ablation experiments are presented in Section 4. The last Section 5 concludes the paper and introduces the future work.

## II. RELATED WORK

### A. Cascade growth prediction

Since the large-scale information cascades with huge destructiveness often burst out suddenly without any warnings [23], how to effectively detect and predict it has become an important topic in the information technology area. A number of studies have investigated the various ways for cascade growth prediction, mainly falling into three categories: the feature-based approaches [5,27,32,33], the generative approaches [8,21,22,29] and the graph representation approaches [3,4,30,31]. Although these methods are very different at the technical level, they all share one similar principle: the predictions are made upon the valuable information captured from an early stage of cascades-for example, Pinto et al. [26] use the early view patterns to predict the popularity of videos on Youtube. Cui et al. [23] detect the outbreak of large cascade according to the users who retweet it at the first stage. Bauckhage et al. [25] predict the future popularity trend of the Internet memes according to initial retweets dynamics. These studies' empirical results indicate a relationship between the early stage of a cascade and its future popularity, especially for the structural and temporal features which are generally effective in different types of information [28]. The question is how to efficiently capture this valuable information and model the relationships between the cascade's early stage and its future popularity.

The regression and classification models [6,28,34] are most commonly adopted to model the growth of cascades based on some well-designed features, such as the content features (e.g., topics [2], hashtags [27], Etc.), the structural features (e.g., node degree [2,24,27], number of border edges [2,24], Etc.) and the temporal features (e.g., time taken for the first retweets [2], number of retweets during a period [24], Etc.). However, the limitations of these feature-based approaches are apparent. First of all, some extracted features are too specific to the particular type of information, making it difficult to be generalized in different scenarios. For example, the hashtag [24] is specially designed for social platforms like Twitter and Weibo. The text sentiment features adopted in [27] cannot be used for photo or video cascades. Moreover, whether the measurement of these selected features can fully capture the predictive information is also questionable. Especially for the structural features that are usually described by several basic indexes (e.g., node degrees [2], number of border edges [24], Etc.) that can hardly reveal the whole topological characteristics of a graph.

Another type of cascade prediction model, the generative approaches, focuses more on temporal dynamics and makes predictions by fitting the time series or macroscopic distributions of retweet into a certain class of functions [8]-for example, Gao et al. [9] build a time relaxation function and a reinforcement function to depict the decay effect and the rich-get-richer phenomenon in cascade growth. Zhao et al. [21] employ the self-exciting point process model to simulate the retweet dynamics of information cascades. However, due to the cascade evolution mechanism remaining elusive, the models have to be built based on strong assumptions and oversimplifications of reality. As a result, there is often a big gap between the model predictions and the actual cascade sizes in real applications.

Recently, as the unsupervised data input and excellent nonlinear modeling capabilities, the deep learning method attracts more attention in the cascade prediction area. Li et al. [3] propose the first end-to-end cascade prediction system (DeepCas) that learns the representation of each cascade graph through a series of node sequences sampled by random walk. In another similar study, instead of using random walk, Cao et al. [4] directly transformed a cascade graph to node sequences following each information diffusion path. Liu et al. [30] apply an attention-based recurrent neural network to deal with the time effect in cascade.

However, all these methods share one similar weakness of standard RNNs or CNNs, which cannot directly handle the non-Euclidean data and rely on random walk to sample many possible paths from the graph. Moreover, it is a tough job to set an appropriate sample size where a small size is not enough for large-scale graphs, but a large sample size can significantly increase the computational redundancy. Therefore, some studies try to develop other methods to learn the graph embeddings without data transformation-for example, Feng et al. [14] calculate the similarity between cascade graphs through the hand-crafted features and use the skip-gram algorithm to make cascades with similar features get similar embeddings. Zhao et al. [31] propose a collaborative embedding framework where the nodes co-occur in the same cascade will obtain similar embeddings. Although the methods above get rid of the data conversion problem, they still need a lot of work to pre-train the embeddings and can only be used in a static network, which is not suitable for real applications in open systems like social network platforms.

Besides, as the graph representation approaches mentioned above either transform the graph as nodes sequences or only focus on the whole embeddings of cascade graphs [3,14], they usually make the models not compatible with nodes' temporal features. Although DeepHawk [4] and T-DeepHwak [35] consider temporal information of cascades, they are not easy to be generalized to different types of cascade.

In this paper, we will build an innovative diffusion-based graph learning framework for cascade prediction, which is different from the current graph representation approaches. The proposed method can not only automatically capture predictive structure information from the cascade graph, but also incorporate the temporal features of each retweet. More importantly, our model is able to directly handle the graph data of information diffusion networks without any data transformation work. The content features are excluded in this paper as they are not general for all types of information and less predictive than structural and temporal features [2].

### B. Graph neural networks

Graph neural networks (GNNs) are motivated by the standard convolutional neural networks (CNNs) that use a shared filter to extract the localized spatial features and compose them to construct highly expressive representations [36]. Similar with the images, the whole representation of a graph can also be obtained by assembling all localized structural features. However, the standard CNNs cannot directly operate on the non-Euclidean data that usually relies on preprocessing work (e.g., random walk) to transform the graph to regular Euclidean data (node sequences). As a result, the conventional graph representation models, like Deepwalk [11], Node2Vec [12], and Struc2Vec [13], lack the ability to deal with dynamic graphs and non-shared parameters lower computation efficiency as well.

Therefore, the GNNs are developed as a generalization of CNNs to graphs and inherit the advantages of CNNs. For example, the GNN proposed in [37] is the first deep learning method that directly processes graph data and makes embeddings for nodes. The rationale of GNN is not too complicated where the hidden state embedding of a node is determined by the information of its neighbours, and the GNN applies a shared local transition function for every node in the graph to aggregate the information from its neighbours. The global embedding of each node will be obtained after several iterations in which the final embedding contains not only information of its neighbours but also the information of neighbours' neighbours. It can be easily seen that the local transition function in GNN plays a similar role with the convolutional filter in standard CNNs.

In addition, as the GNN makes a huge success in graph learning, many variants of GNN are developed in the following studies. For example, the DGP [38] extends the GNN for directed graphs; the G2S [39] incorporates the edge information into the model; the GGNN [15] combines GRU with the update process of node hidden state. In general, the current GNNs can be divided into two categories according to convolution approaches: the spectra methods and non-spectra methods. In a spectra method, the convolution operation is defined in the Fourier domain by computing the eigendecomposition of the graph Laplacian, for example, the GCN [40]. In contrast, the non-spectra method aggregates information directly from a node's local neighbourhood, such as GraphSage [16]. No matter which type of convolution methods are adopted, numerous empirical results show that the GNNs are more efficient and effective in capturing the structural features of graphs than conventional graph representation methods. In our study, we are going to extend the GNNs to the information cascade domain and propose a graph convolutional learning framework for cascade prediction, which has not been shown in other studies. Besides, we use a non-spectral convolution approach to handle the directed cascade graphs.

### III. METHOD

As the final cascade grows from its historical diffusion graph, the future popularity depends on the features of nodes that have already been in the cascade [2], i.e., the structural and temporal features in this study. In reality, it is easy to know who retweets a message or cites an academic paper, and when it happens. Then, an early stage of a cascade graph can be obtained to predict future popularity. In this section, we first define the problem and introduce how our method extracts the features of cascades and makes predictions.

### A. Problem Definition

An accurate prediction of future cascade sizes is of paramount importance for many real applications like rumor prevention, hot topic detection, viral marketing, and so on. In this study, instead of working with binary prediction that determines whether an information cascade will be popular or not, we predict the exact growth size (i.e., popularity) of a cascade after a certain period, according to its early information diffusion graph.

**Definition 1 (Global network)** Suppose that at time $t_0$ we make a snapshot of a social network $G = (V, E)$, where $V$ is the set of nodes in this network at $t_0$ and $E \subset V \times V$ is the set of edges between nodes. For example, a node can be a user in a social platform or a paper in a citation network. An edge shows the relationship between two nodes.

**Definition 2 (Cascade graph)** We denote a set of information cascades in the global social network $G$ as $C$. Each cascade $c \in C$ with a duration $t$ after its origination is described by a directed graph $g_c^t = (V_c^t, E_c^t, T_c^t)$, where $V_c^t$ is a set of nodes that have got involved in the cascade $c$ within duration $t$ after the original post (e.g. the users who retweet the message or the authors who cite the paper), an directed edge $(v_i, v_j) \in E_c^t$ represents that node $v_j$ adopts the cascade from node $v_j$, and a time label $t_v \in T_c^t$ denotes the time elapsed between the original post and node $v$'s retweet.

**Definition 3 (Growth size)** In this study, the cascade size is defined as the number of retweets or citations of a message or academic paper. Due to the rich-get-richer phenomenon, there is usually an intrinsic correlation between the observed cascade size and its final size. In order to exclude its impact on our model, we follow the designs of previous works [3,4,14,30] that predict the increment of a cascade's size after a given time interval $\Delta t$ instead of directly predicting its final size. The growth size can be denoted as $\Delta s_c = |V_c^{t+\Delta t}| - |V_c^t|$, and the $|V_c^{t+\Delta t}|$ is closer to the final size of a cascade when $\Delta t$ is larger.

According to the framework of our model shown in Fig. 1, we take the cascade graph of certain time $t$ as input and output is the predicted increment of cascade size at time $t+\Delta t$. The model automatically embeds the structural features of nodes into a matrix through the proposed graph convolutional layer and combines each node's temporal features for the final prediction. Let $\mathbf{H}_c$ and $\mathbf{T}_c$ be the matrix and vector respectively constructed by stacking all the obtained structure representations and temporal features of nodes in graph $g_c^t$. Then, the cascade prediction can be formulated as, given $t, \Delta t$, and $\{g_c^t, \Delta s_c\}_{c \in C}$, finding the optimal mapping function $f$ that minimizes the following objective function:

$$O = \frac{1}{|C|} \sum_c (f(\mathbf{H}_c, \mathbf{T}_c) - \Delta s_c)^2 \quad (1)$$

### B. Model
**Node embedding**
In order to input a cascade graph into the graph convolutional layer, we first generate an initial embedding for each node in the global graph G. Each node is represented as an one-hot vector $q \in \mathbf{R}^{N_{node}}$, where $N_{node}$ is the number of nodes. All nodes share an embedding matrix $A \in \mathbf{R}^{H \times N_{node}}$ that transforms each node into its initial embedding vector $x = Aq$, $x \in \mathbf{R}^H$.

**Graph convolutional layer**
The increment of cascade size can be regarded as the consequence of information diffusion graph growth, as shown in Fig. 2. The number of new nodes to occur in the cascade depends on the features of 'propagators', i.e., the nodes in the current diffusion graph.

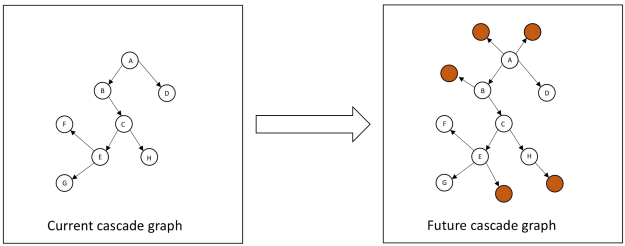

**FIGURE 2.** The growth of cascade graph

The first component of our model is the graph convolutional layer that learns node representations containing localized structural features of each node. In reality, the whole topological graph of a social network is usually not observable and too large to learn. Therefore, as a subgraph of the social network, the cascade graph containing abundant structural information is often used to train the node embeddings instead of the whole social network [41]. In our work, we also obtain a node's structural features according to the cascade graph involving this node.

*Aggregator function.* As the localized structure of a node is determined by its neighbours, the first step of the graph convolutional layer is going to aggregate the information of each node's local neighbours. A fixed-shape filter is not suitable for graph learning because a graph is non-Euclidean where the number of each node's neighbours is not fixed and the order of neighbours has no effect on its structure. Therefore, we adopt a non-spectra convolution method to aggregate the localized structural features of each node. Additionally, as the aggregator function should be symmetric (i.e., invariant for the input order of nodes) and trainable for maintaining high representational capacity, a sum aggregator combined with a gated update is adopted in our model.

*Aggregation depth.* We also make $k$ iterations of aggregation function for each node to capture the structural features of neighbours in deep layers. Fig. 3 clearly illustrates how the process works, when $k=1$ the node $C$ embeds the information of its local neighbours into its hidden state $\mathbf{h}_C^1$, meantime the updated $\mathbf{h}_B^1$ of node $B$ also includes information from its neighbour, i.e., node $A$ which is one of the second layer neighbours of node $C$. After that, the information of node $A$ that is embedded in $\mathbf{h}_B^1$ will be captured by node $C$ in the next iteration. As a result, after $k$ iterations, the obtained node $v$'s hidden state $\mathbf{h}_v^k$ contains the information from its $k_{th}$ layer neighbours.

*Bio-directional aggregation.* Taking into account the information diffusion directions, we develop a bio-directional aggregation approach. As shown in Fig. 3, when the model extracts structural features of node $C$, gray arrows and blue arrows respectively represent that the model aggregates information of incoming neighbours and outgoing neighbours. As the propagation of a cascade is always a one-way process, distinguishing nodes from different directions brings more information of structural features. For example, in Fig. 3, if we treat the cascade graph as undirected, the model will capture the information that the message can also spread from node $D$ to node $C$, which is not in line with the reality. An example of this is that a paper can never cite another paper published after it. Therefore, according to the cascade graph, we generate two adjacent matrices **A_in** and **A_out** that respectively record each node's neighbours from two directions. Then, the convolutional layer can be represented as follows:

First, each node gets an initial embedding $\mathbf{H}^0 = \mathbf{X}$, where $\mathbf{H}^k \in R^{|V| \times D}$ and $\mathbf{X} \in R^{|V| \times D}$ are respectively the stacked $k_{th}$ hidden state and the initial embedding of each node, $|V|$ is the number of nodes in the cascade graph, and $D$ is dimension size of node embedding. Then the bio-directional aggregation of node $v$'s localized structural features is represented as:

$$\mathbf{h}^k_{N(v)} = concat(\mathbf{a}^v_{in} \cdot \mathbf{H}^{k-1}, \mathbf{a}^v_{out} \cdot \mathbf{H}^{k-1}) \quad (2)$$

Where $\mathbf{h}^k_{N(v)} \in R^D$ is the aggregated hidden states of node $v$'s neighbours at $k_{th}$ iteration, $\mathbf{a}^v_{in} \in R^{|V|}$ and $\mathbf{a}^v_{out} \in R^{|V|}$ are two rows of adjacent matrices **A_in** and **A_out** corresponding to the node $v$, · is the matrix multiply that makes the sum aggregation of node $v$'s localized structural features.

*Gated update.* As the convolution layers comprise $k$ iterations of aggregator function, in our work we use a GRU(Gated Recurrent Unit) to update the hidden state of each node. In specific, the reset gate $\mathbf{r}^k_v \in R^D$ of node $v$ is computed as:

$$\mathbf{r}^k_v = \sigma(\mathbf{W}_r \cdot \mathbf{h}^k_{N(v)} + \mathbf{U}_r \cdot \mathbf{h}^{k-1}_v) \quad (3)$$

Where $\sigma$ is the sigmoid activation function, $\mathbf{W}_r \in R^{D \times D}$ and $\mathbf{U}_r \in R^{D \times D}$ are parameters learned during training. The update gate $\mathbf{z}^k_v \in R^D$ of node $v$ is shown as (4) where $\mathbf{W}_z \in R^{D \times D}$ and $\mathbf{U}_z \in R^{D \times D}$ are also trainable parameters.

$$\mathbf{z}^k_v = \sigma(\mathbf{W}_z \cdot \mathbf{h}^k_{N(v)} + \mathbf{U}_z \cdot \mathbf{h}^{k-1}_v) \quad (4)$$

After that, the final updated hidden state of node $v$ is computed as:

$$\mathbf{h}^k_v = (1 - \mathbf{z}^k_v) \otimes \mathbf{h}^{k-1}_v + \mathbf{z}^k_v \otimes \widetilde{\mathbf{h}}^k_v \quad (5)$$

Where $\widetilde{\mathbf{h}}^k_v = \tanh(\mathbf{W} \cdot \mathbf{h}^k_{N(v)} + \mathbf{U} \cdot (\mathbf{r}^k_v \otimes \mathbf{h}^{k-1}_v))$, $\otimes$ is the element-wise multiply between vectors, $\mathbf{W} \in R^{D \times D}$, and $\mathbf{U} \in R^{D \times D}$. Let $\mathbf{H}_c$ be the matrix constructed by stacking all the obtained representations of nodes in graph $g_c$, after $k$ iterations, the $\mathbf{H}_c$ is computed as:

$$\mathbf{H}_c = (\mathbf{h}^k_1, \mathbf{h}^k_2, ..., \mathbf{h}^k_{|V|})^T \quad (6)$$

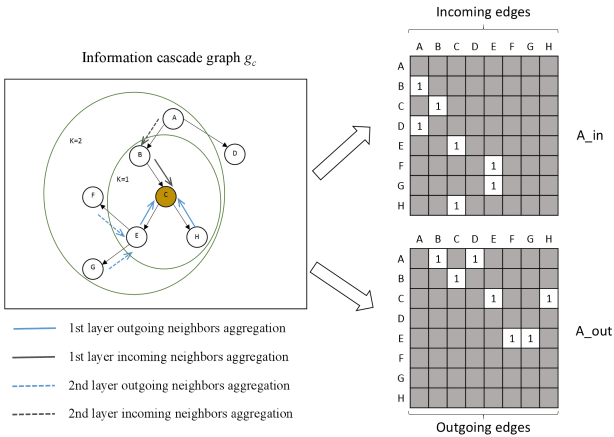

**FIGURE 3.** Bio-directional aggregation

**Temporal features extraction**

Time effect is a common phenomenon in information cascade diffusion and cannot be ignored in cascade prediction. For instance, twitter is often frequently retweeted during the first period after it is posted, and the number of new retweets declines with time. However, the end-to-end models learn cascade representations on path-level or graph-level, making it difficult to incorporate the temporal features of each node. In our work, the graph convolutional layer outputs nodes' representations that are also compatible with the node's temporal features.

Assuming a cascade graph $g^t_c = (V^t_c, E^t_c, T^t_c)$ is observed with a time duration $t$ after its origination, it is easy to know how long the time eclipses between its origination and each retweet/citation. For example, each micro-blog records its post time on the Weibo platform, once another user retweet it, we can also see the time when it happens from the retweet list. Then we can obtain a set of time labels $\{t_v = t^r_v - t^c_0\}(0 \leq t_v \leq t, v \in V^t_c)$ of each node in cascade graph $g^t_c$, where $t^r_v$ is the time when node $v$ retweets the message and $t^c_0$ is its original post time. A normalization is made to all time labels as shown in (7). Then we stack the normalized time labels into the vector $\mathbf{T}_c$ that is regarded as the temporal features of nodes in cascade graph $g^t_c$.

$$t'_v = t_v / t \quad (0 \leq t_v \leq t) \quad (7)$$

$$\mathbf{T}_c = [t'_1, t'_2 ... t'_{|V|}]^T \quad (8)$$

**Attention-based aggregation layer**

As the given cascade graph $g^t_c = (V^t_c, E^t_c, T^t_c)$ shown in Fig. 1, through the first part of our model we obtain the matrix $\mathbf{H}_c$ and vector $\mathbf{T}_c$ that respectively contain the structural features and temporal features of each node. It is obvious that the information contained in $\mathbf{H}_c$ and $\mathbf{T}_c$ is redundant for final prediction, since the nodes with different influence usually provide different contributions for cascade growth in reality. For example, in Fig. 2, not every node in the cascade graph brings new retweet. In our work, it is the key to identify the useful features in $\mathbf{H}_c$ and $\mathbf{T}_c$, and abandon those useless information for prediction. Therefore, we build another aggregation layer which is inspired by the attention mechanism to automatically select useful features.

$$g(\mathbf{H}_c, \mathbf{T}_c) = relu\left(\sum_{v \in V^t_c} \left(\sigma(i(\mathbf{H}_c, \mathbf{T}_c)) \otimes \tanh(j(\mathbf{H}_c, \mathbf{T}_c))\right)\right) \quad (9)$$

The calculation of the aggregation layer is shown in (9), where $\otimes$ is an element-wise product, $\sigma(i(\mathbf{H}_c, \mathbf{T}_c))$ acts as a soft attention mechanism that identify which node's feature is more important, $i$ and $j$ are neural networks that take the concatenation of $\mathbf{H}_c$ and $\mathbf{T}_c$ as input and outputs real-valued vectors. It can be clearly seen that the learned cascade representation $g(\mathbf{H}_c, \mathbf{T}_c)$ contains the best information for future growth prediction.

**Output layer**

The output layer of our model is a fully connected neural network, taking the learned cascade representation $g(\mathbf{H}_c, \mathbf{T}_c)$ as input and outputting the final prediction of growth size: $f(\mathbf{H}_c, \mathbf{T}_c) = \mathrm{MLP}(g(\mathbf{H}_c, \mathbf{T}_c))$, where MLP stands for a multi-layer perception. In the end, the eventual objective function to be minimized is defined as:

$$O = \frac{1}{|C|} \sum_c (f(\mathbf{H}_c, \mathbf{T}_c) - \Delta s_c)^2 \quad (10)$$

where $f(\mathbf{H}_c, \mathbf{T}_c)$ is the predicted increment size of cascade $c$, $\Delta s_c$ is the cascade $c$'s actual growth size, $|C|$ is the total number of information cascades.

## IV. EXPERIMENT

We apply our model to real information cascades phenomena in social network and academic paper citation to evaluate the performance of our model. We also compare the performances of our model with other state-of-art cascade prediction methods to illustrate the advantages of CasGCN. In addition, we make several variants of our CasGCN model to test the effectiveness of aggregator function and temporal features incorporated in our model.

### A. Dataset
**Sina Weibo**

The first application scenario of our method is to predict the retweet popularity in Sina Weibo, which is one of the largest social platforms in the world. Weibo has more than 500 million monthly active users, and nearly 10 million original micro-blogs are posted every day. These messages can spread by users' retweet behaviors, and some of them can eventually become influential information cascades on the Weibo platform. In this work, we generate the diffusion graph for each micro-blog according to the information provided by Weibo.cn. For example, on the page of a micro-blog at Weibo.cn, we can obtain all the retweet records from the retweet list, where "//@B//@A: ..." shows the user B retweet this message from the user A, and "C: ..." means the user C retweet it directly from the original poster. Thus, we can know not only which nodes but also how and when they get involved in this information cascade.

In this work, we use the Selenium (a crawler tool of python) to crawl the retweet information of popular micro-blogs from February 2020 to June 2020. Since the temporal dynamics of retweet are quite different between day and night, the micro-blogs selected in this paper are all posted during 8:00 AM to 18:00 PM. As we mentioned in problem definition, based on the observed cascade graph with a time duration $t$ after origination, our model predicts the increment size of a cascade after a given time interval $\Delta t$. Following previous literature[4], we set the observation time $t=3\ hours$ and $\Delta t = 21\ hours$ for Weibo data. In other words, our model predicts the popularity of a micro-blog within 24 hours after origination. Therefore, the $\Delta s_c$, i.e. the increment size of a cascade, can be computed as $\Delta s_c = R_c^{24} - R_c^{3}$ where R is the number of retweets.

Besides, we notice that a majority of micro-blogs are with small observed cascade sizes and have little growth in the end, which can significantly influence the overall performance of a model. In other words, the model may tend to make small predictions for better overall performance, since only a few cascades in the data set make a large growth and their prediction errors have little impact on the total performance. For example, the datasets adopted in [3,4,14] are extremely imbalanced where the most tweets or micro-blogs make zero growth after $\Delta t$. As a result, although the overall performances of those models are good, they may fail to learn the patterns of those large cascades which are much more important in cascade predictions. In contrast with those studies, we try to focus more on the model's ability of accurately predicting the increment of those large cascades. Therefore, we use the crawled Weibo data to construct 3 data sets, i.e. W1, W2 and W3. Specifically, the data set W1 selects the micro-blogs with more than 30 nodes in the observed cascade graphs. The data set W2 and W3 respectively consists of the micro-blogs with more than 60 and 90 observed nodes. The distributions of 3 data sets are shown in Fig. 4, where it can be clearly seen that the number of cascades with small growth in W3 and W2 is less than W1.

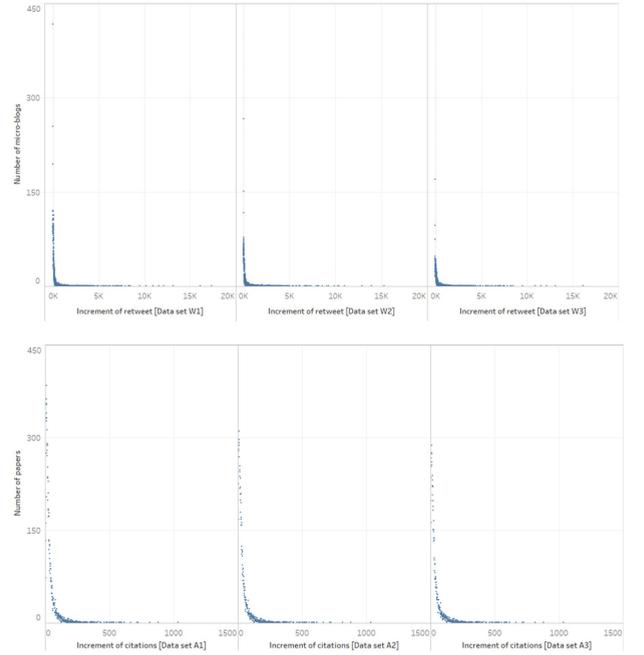

**FIGURE 4.** Distribution of cascade growth sizes

**Academic paper citation**

Another application of our model is to predict academic paper citations. The data set adopted in this paper is the **DBLP-Citation-network V9** that contains 3,680,007 papers and 1,876,067 citation relationships until 2017. We treat each paper as a node and construct the cascade graph according to its citation network. For example, in Fig. 5, given a target paper A to be the root, we first find all the nodes that cited this paper within the observation time t and connect them with A by directed edges. Meantime we check the citation relationships between A's neighbours. A directed edge will also be built to connect B and C if B cited C. We set the prediction time interval $\Delta t$ =15 years and the observation time $t$=5 years after publication in this paper. Then, we also use the papers published between 1987 and 1997 to generate 3 data sets, i.e. A1, A2 and A3 that respectively contain the papers with more than 10, 20 and 30 nodes in the observed cascade graphs. The statistics of our data sets used in this paper is reported in Table I.

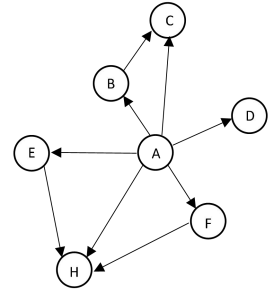

FIGURE 5. Paper citation network

TABLE I
STATISTICS OF DATA SETS

| | | Weibo | | | Aminer | | |
|---|---|---|---|---|---|---|---|
| Data set | | W1 | W2 | W3 | A1 | A2 | A3 |
| Number of nodes in the observed cascade graph | | N>30 | N>60 | N>90 | N>10 | N>20 | N>30 |
| Number of cascades | Train | 22427 | 16357 | 12165 | 18652 | 12834 | 9487 |
| | Val | 4482 | 3270 | 2433 | 3730 | 2566 | 1897 |
| | Test | 4483 | 3270 | 2433 | 3730 | 2567 | 1898 |
| Avg. Number of leaf nodes per graph | Train | 297.8 | 382.7 | 487.2 | 27.9 | 40.5 | 49.6 |
| | Val | 278.2 | 354.4 | 481.2 | 19.6 | 36.3 | 40.8 |
| | Test | 288.5 | 386.5 | 490.3 | 23.5 | 33.2 | 45.9 |
| Avg. Number of edges per graph | Train | 476.5 | 760.4 | 902.6 | 67.8 | 98..4 | 113.8 |
| | Val | 417.0 | 591.6 | 877.3 | 43.9 | 88.6 | 99.2 |
| | Test | 490.5 | 695.7 | 863.5 | 61.1 | 82.3 | 103.1 |

### B. Evaluation metric

The mean squared log-transformed error (MSLE), an effective indicator to measure the difference between predicted values and actual values, is adopted as the evaluation metric in this paper. The MSLE is a variant of MSE (mean squared error), and it is frequently applied in regression tasks and cascade prediction studies [3,4,14]. The definition of MSLE is shown as following:

$$MSLE = \frac{1}{|C|}\sum_{c=1}^{|C|} SLE_c^2 \quad (11)$$

where $|C|$ is the total number of cascades, an arbitrary cascade $c$'s squared log-transformed error $SLE_c = (\log(\Delta s'_c +1) - \log(\Delta s_c +1))^2$, $\Delta s'_c$ is the predicted increment of cascade size, and $\Delta s_c$ is the actual growth size of the cascade.

### C. Baseline methods

In this paper, we also adopt several common and state-of-art cascade prediction approaches as the baselines to compare with the proposed CasGCN, including the feature-based methods, the DeepCas, the DeepHawk, and the struc2vec graph embedding method.

**Feature-linear and Feature-deep.** The linear regression is one of the most common approaches used to model the relationship between cascade popularity and the hand-crafted features in the cascade prediction area [5,27,34]. Except for this, we also use a neural network to combine the selected features with cascade growth size in a non-linear model. In this paper, we extract several frequently-used structural and temporal features that can be generalized across all the data sets in this study.

*Structural features.* Since the Weibo platform shuts down the data interface of users' followers, we can only obtain the structure information of cascade graphs. We first count the number of leaf nodes in each graph [2,5], we also calculate the average and the max degree of nodes [28], and the average and max path length [24] are included to measure the depth of each graph.

*Temporal features* of nodes can be used to evaluate information diffusion speed, which are proved as important as structural features for cascade prediction [2]. We consider the average time elapsed between the message origination and each retweet/citation [24], the average and the max time interval between two successive retweets/citations [22].

We calculate the values of these selected features for each observed cascade graph in our data sets. Then, the obtained features vectors are fed to both the linear regression model and fully connected neural network to learn to estimate the increment of cascade sizes.

**DeepCas** is one of the state-of-art deep learning methods in the cascade prediction area. The DeepCas is an end-to-end deep learning framework which transforms cascade graphs into a series of node sequences through random walk and feeds the sequences into a deep learning network to predict cascades growth size. In this paper, we adopt it as a baseline method and set the number of walk K = 200 sequences with walk length T = 10 for random walk, following the existing studies [3].

**DeepHawk** is a combination of deep learning and generative approach (i.e. the Hawk process), which makes each information diffusion path as an input of a recurrent neural network. Then, a learned time decay effect is added to each path which is used to predict the increment size of a cascade through MLP. This approach mainly considers the temporal dynamics of cascade growth instead of the effect of cascade graph structure.

**Struc2Vec** is one of the strongest node embedding methods which can effectively capture each node's structural features and embed it into a vector [13]. The Struc2Vec constructs a new graph according to each node's degrees of different layers and uses the skip-gram to make the nodes with similar structure obtain similar embeddings. As a baseline in this study, we first use Sturc2Vec to learn the embedding of each node in a cascade graph and feed them into a recurrent neural network with an order of retweet/citation time. The final prediction of cascade growth size is obtained through MLP.

### D. Performance Comparison

We apply our proposed CasGCN model and five baseline methods on six data sets of two cascade prediction scenarios, i.e., the retweet popularity of Weibo and academic papers' citations. In this section, we are going to compare the performances of all models. The result statistics are shown in Table II. The differences between three data sets in the same scenario are scales of selected cascades graphs, by which we can better investigate the models' prediction abilities on different-scale cascades.

It can be clearly seen from Fig. 6 that the overall performances of all models degrade with the growing cascade scales selected in data sets. Besides, compared with predicting Weibo retweet popularity, all the models obtain lower MSLEs on predicting paper citations in which the cascade sizes are often smaller than Weibo retweet popularity. The results show that accurate predictions for large cascade growth are much more difficult than for small ones.

It can also be found that our proposed model outperforms all the baseline models in predicting both retweet popularity and paper citations. Note that for the data sets with less small cascades, our proposed model reaches significantly better results than the feature-based models, the DeepCas and the DeepHawk. It fully illustrates the advantage of our model in handling large and complex cascade graphs.

Surprisingly, the performances of the Struc2Vec model are close to our proposed model, which makes predictions based on the trained node embeddings. Compared with the random walk and hand-crafted features, the graph convolutional neural network of our model and the Struc2Vec seems to be more effective in capturing predictive structural features of large-scale cascade graphs. However, the training process of Struc2Vec is extremely time-consuming, where we spend nearly two days on training node embeddings with a 2.4GHz CPU, 120G RAM, and a GTX 950M GPU. In contrast, a training epoch of our proposed model only costs about 10 minutes.

For two feature-based models in this study, the feature-deep model outperforms the feature-linear model in both scenarios. The result is similar to [4], which incorporates nodes' temporal features into the models as well as this paper. In contrast, two feature-based models obtain approximate performances in experiments of [3], where they only extract nodes' identities and structural features. The results imply that non-linear relationships exist between cascade increment sizes and nodes temporal features.

We can also see that the DeepCas, one of the state-of-art cascade prediction models, outperforms the Feature-based model in paper citation predictions but obtains worse performances in predicting Weibo retweet popularity. It proves that the DeepCas's prediction ability in large-scale cascades can be limited by random walk parameters (i.e., number of walks and walk length). Comparing the results of DeepCas with DeepHawk, we find that the DeepHawk, based on information diffusion paths and time decay effect performs better than the whole end-to-end DeepCas model. Besides, the training time of the DeepCas model and DeepHawk model are almost twice that of the CasGCN model.

Overall, our proposed CasGCN model shows a strong ability to extract predictive structural features of cascade graphs, and it obtains superior performances in both retweet popularity and paper citation predictions. Our proposed model performs significantly better than not only the feature-based models but also the state-of-art deep learning models, especially for data sets with more large-scale cascades.

TABLE II
OVERALL PREDICTION PERFORMANCE

| | Weibo | | | Aminer | | |
|---|---|---|---|---|---|---|
| Data set | W1 | W2 | W3 | A1 | A2 | A3 |
| Number of nodes in observed cascade graph | N>30 | N>60 | N>90 | N>10 | N>20 | N>30 |
| Feature-linear | 3.267*** | 3.594*** | 3.768*** | 1.695*** | 1.896*** | 2.141*** |
| Feature-Deep | 2.632*** | 2.886*** | 3.423*** | 1.536*** | 1.884*** | 2.071*** |
| DeepCas | 2.753*** | 2.975*** | 3.578*** | 1.283*** | 1.465*** | 1.877*** |
| DeepHawk | 2.152*** | 2.415*** | 2.987*** | 0.967*** | 1.187*** | 1.528*** |
| Struc2Vec | 1.963** | 2.238*** | 2.406 | 0.824*** | 1.117** | 1.395* |
| CasGCN (Proposed) | 1.846 | 2.067 | 2.384 | 0.738 | 1.052 | 1.338 |

\*\*\*, \*\*, and \* respectively means the result is significantly different with the proposed model at 0.01, 0.05 and 0.1 level.

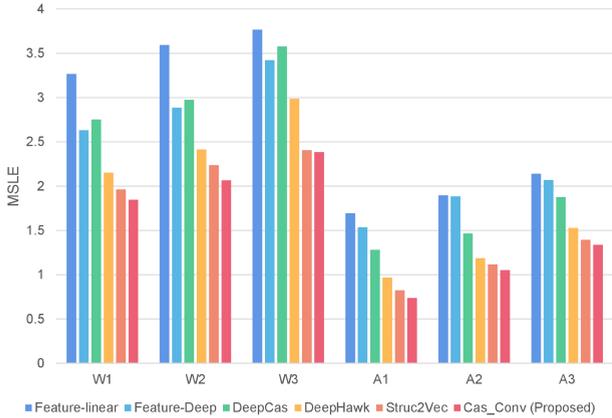

**FIGURE 6.** Comparison of mean squared log-transformed errors

### E. Variants of CasGCN

In order to further evaluate the effectiveness of components in our proposed model, we make several variants of CasGCN. In addition, we also apply these modified models on the above six data sets and compare the results with original CasGCN.

**CasGCN-max and CasGCN-mean**. The aggregator function is one of the most important components in our model that extracts each node's structural features. In CasGCN, we aggregate information from a node's neighbours by summing up all the neighbours' vectors and update the node representation through a gated recurrent unit. On the other hand, max-pooling [16] and mean-pooling [15] are two widely used aggregator functions in the GNN area, which perform well in various application scenarios. Therefore, to test the effectiveness of our proposed aggregator function, we build the CasGCN-max and the CasGCN-mean that respectively use max-pooling and mean-pooling as aggregator functions instead of our proposed method.

**CasGCN-undirected** treats cascade graphs as undirected graphs and does not distinguish a node's neighbours from different information diffusion directions. As a result, the undirected CasGCN model extracts a node's structural features by only one operation of aggregating all its neighbours. We construct this model to test the effectiveness of the bio-directional aggregation proposed in this study.

**CasGCN(no time effect)** is a variant of CasGCN without considering nodes' temporal features and makes predictions only based on the representations obtained from the graph convolutional layer. This model is constructed to test if it is necessary to consider the time effect in the CasGCN model.

TABLE III
PREDICTION PERFORMANCE OF CasGCN VARIANTS

| Data set | Weibo | | | Aminer | | |
|---|---|---|---|---|---|---|
| | W1 | W2 | W3 | A1 | A2 | A3 |
| CasGCN-max | 2.014*** | 2.237*** | 2.638*** | 1.067*** | 1.284*** | 1.531*** |
| CasGCN-mean | 1.907* | 2.104** | 2.479*** | 0.764 | 1.128* | 1.439*** |
| CasGCN-undirected | 2.259*** | 2.439*** | 2.839*** | 0.976*** | 1.318*** | 1.648*** |
| CasGCN(no time effect) | 2.094*** | 2.382*** | 2.549*** | 1.178*** | 1.296*** | 1.568*** |
| CasGCN (Proposed) | 1.846 | 2.067 | 2.384 | 0.738 | 1.052 | 1.338 |

***, **, and * respectively means the result is significantly different with the proposed model at 0.01,0.05 and 0.1 level.

The statistical results of all modified models are shown in Table III. We can clearly see a significant performance reduction of the CasGCN-max compared with the original CasGCN, demonstrating that max-pooling is less effective than the proposed aggregator function in extracting nodes' localized structural features. The CasGCN-mean performs closely to the original CasGCN except for data set W3 and A3, which shows that our gated aggregator function is more suitable than mean-pooling for large-scale cascades.

Additionally, compared with CasGCN-undirected, CasGCN with a bio-directional aggregation reaches remarkably better overall performances. The results indicate that edge directions can provide more useful information about cascade structure, and it is better to distinguish different information diffusion directions in cascade prediction.

Moreover, it can also be seen that omitting the time effect leads to a significant increase of prediction errors where the CasGCN(no time effect) does not perform as well as the original CasGCN on all data sets. It proves that temporal features contain important information for cascade prediction, which cannot be obtained from cascade structures.

In summary, the components of gated aggregator function and bio-directional aggregation in the proposed CasGCN model effectively improve the prediction performances by extracting more valuable information of cascade structures. Besides, the temporal dynamics of cascade growth are as important as structural features for future cascade prediction, and it is necessary to consider the time effect to improve prediction performances. The experimental results demonstrate the effectiveness and necessity of all three components in the CasGCN model.

### V. DISCUSSION AND CONCLUSION

This study presents a graph convolutional based deep learning framework to predict future cascade growth based on topological structures and temporal dynamics of nodes in a cascade graph. Through a carefully designed non-spectra graph convolutional layer, the model extracts nodes' localized structural features along both information diffusion directions. This is followed by applying an attention-based aggregation layer to combine the extracted feature from graph convolutional layer with nodes' temporal features to make predictions. As a result, the model not only inherits a strong ability of graph neural network (GNN) in handling complex graph structures but also improves training efficiency as there is no data transformation process. Besides, the compatibility with other node-level features allows the model to conveniently optimize performance by incorporating predictive information (e.g., temporal features) that cascade structure cannot provide. The experiments are done on two information cascade scenarios, i.e., predicting Weibo retweet popularity and predicting academic paper citations, in which our proposed CasGCN outperforms feature-based methods, the state-of-art DeepCas, DeepHawk, and powerful node embedding method Struc2Vec.

This study makes a successful application of GNN on cascade growth prediction where the proposed graph convolutional layer (i.e., gated aggregator function and bio-directional aggregation) show strong ability in extracting structural features of cascade diffusion graphs. The outcomes provide new evidence of GNN's great potential in dealing with social network and information diffusion problems.

In addition to these, we have two other interesting findings from our empirical experiments. First, previous results can hardly reveal a model's capacity on dealing with large-scale cascades, as most information cascades are with small future growth in data sets obtained from the reality. In other words, since the number of small cascades is far more than that of large cascades, it is easy for a model to reach a good overall performance by making smaller predictions. In this study, we reduce the number of small cascades by selecting cascades with more nodes at observation to make the obtained results better tell if the model learns the patterns of large-scale cascades. Since predictions of large-scale cascades are much more important, the test of models should pay more attention to performances on large-scale cascades instead of overall performances in future cascade prediction studies.

Another interesting finding is the non-linear relationships between nodes' temporal features and cascade growth sizes, which limits the performances of linear models in cascade prediction. This finding is contrary to previous studies [3,14], which have suggested that as long as a sufficient number of right features are selected, feature-based linear models can reach similar performances as end-to-end models. On the one hand, factors that affect information diffusion are still inconclusive, and it is not easy to include all useful features in a linear model. On the other hand, linear models are unable to handle the non-linear effects (e.g., time effect) of factors sufficiently. Although a deep learning model is not as interpretable as a linear model, it still has the advantages of capturing those unknown and complex cascade evolution mechanisms.

In future works, we will consider optimizing the CasGCN model from two aspects. One is to analyze the learned representations of cascades and develop a method that can better interpret how nodes' structures and temporal dynamics affect cascade growth. The other is to incorporate other features (e.g., content features, user features, Etc.) into the model to find possible correlations between those features and cascade growth sizes.

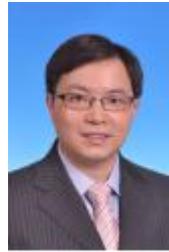

**Minghui Qian** is an Associate Professor and Doctoral Supervisor at the School of Information Resource Management, Renmin University of China. He is the Head of the Information Resource Management group and an Associate Dean of the Marketing Research Centre of China at Renmin University of China. He is also the Associate Dean of Brand Specialty Committee of Chinese Business History Society, the member of Agricultural Product Processing Industry Expert Board of Ministry of Agriculture (MOA) of the People's Republic of China, the evaluation expert of Chinese Postdoctoral Science Foundation, and the evaluation expert of Academic Degree & Graduate Education Development Centre of Ministry of Education (MOE) of the People's Republic of China.

His main research areas include brand management, information analysis, and information resource management. He has published more than 100 research articles in Chinese and foreign journals, 4 monographs, 7 textbooks, and 8 research reports. He has involved in various projects including the Program of Humanity and Social Science Youth Foundation of the Ministry of Education of China, National Natural Science Foundation of China and a project of Beijing Committee of Science and Technology in China.

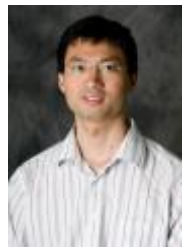

**Xiaowei Huang** (PhD of Computer Science, Chinese Academy of Science), is a reader in the Department of Computer Science at the University of Liverpool, U.K.
(email: xiaowei.huang@liverpool.ac.uk).
His research is in building the robustness and resilience of artificial intelligence.

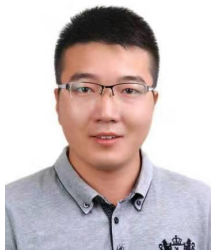

**Zhixuan. Xu** received the B.S. degree in communication engineering from Tongji University, Shanghai, China, in 2015 and the M.S. degree in informatics from Renmin University, Beijing, China, in 2017. He is currently pursuing the Ph.D. degree in data analysis at Renmin University, School of information resource management, Beijing, China.

From 2019 to 2020, he was a visiting researcher with the Department of Computer Science, University of Liverpool, Liverpool, UK. His research interest includes the development and application of machine learning and graph deep learning algorithms in processing real problems in information diffusion, information retrieval, and information recommendation.

Mr. Xu's awards and honors include the Outstanding Paper Award of 2018 Information Science Doctoral Forum of China, 2018 JMS China Marketing Science Annual Conference Excellent Paper Award.

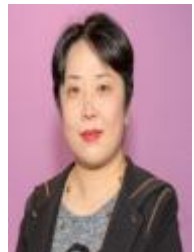

**Jie Meng** is a lecturer in the Institute of Digital Technologies at Loughborough University London, U.K.
(email: j.meng@lboro.ac.uk).

Jie has a PhD in Marketing from the University of New South Wales, Australia, and a GCETT certificate of Higher Education from Federation University, Australia. Before joining Loughborough University London, Jie expanded her footprint as a digital marketing lecturer and researcher in Asia, Australia, and the UK. Her research involves social media and social network, digital analytics and digital wellbeing.